\documentclass[12pt]{elsarticle}

\usepackage{graphicx} %enable figures
\usepackage{multirow} %enable `multirow' capability in tables
\usepackage[lmargin=1.25in,tmargin=1.25in,bmargin=1.25in,rmargin=1.25in]{geometry}
\usepackage[colorlinks=true, allcolors=blue]{hyperref} %reference and citation hyperlinking linking
\usepackage{url}
\usepackage{amsmath}
\usepackage{appendix}
\usepackage[normalem]{ulem}

\newcommand{\fig}[3]{\begin{figure}[h]
\centering\includegraphics[width=0.5\textwidth]{#1}\caption{#2}\label{#3}
\end{figure}}

\newcommand{\red}[1]{#1}%{\textcolor{red}{#1}}
\renewcommand{\deg}{^\circ}
\newcommand{\Uttf}{^{235}\mathrm{U}}
\newcommand{\Utte}{^{238}\mathrm{U}}

\begin{document}

%Define fields for \maketitle

\title{Neutron diagnostics for the physics of a high-field, compact, $Q\geq1$ tokamak}

\author{R.A.~Tinguely\fnref{fn1}\corref{cor1}}
\author{A.~Rosenthal\fnref{fn1}}
\author{R.~Simpson\fnref{fn1}}
\author{S.B.~Ballinger}
\author{A.J.~Creely}
\author{S.~Frank}
\author{A.Q.~Kuang}
\author{B.L.~Linehan}
\author{W.~McCarthy}
\author{L.M.~Milanese}
\author{K.J.~Montes}
\author{T.~Mouratidis}
\author{J.F.~Picard}
\author{P.~Rodriguez-Fernandez}
\author{A.J.~ Sandberg}
\author{F.~Sciortino}
\author{E.A.~Tolman}
\author{M.~Zhou}
\author{B.N.~Sorbom}
\author{Z.S.~Hartwig}
\author{A.E.~White}

\address{MIT Plasma Science and Fusion Center, Cambridge, MA, USA 02139}

\fntext[fn1]{Co-first-authorship}
\cortext[cor1]{Author to whom correspondence should be addressed: rating@mit.edu}

\begin{abstract}

\red{Advancements in high temperature superconducting technology have opened a path toward high-field, compact fusion devices. This new parameter space introduces both opportunities and challenges for diagnosis of the plasma. This paper presents a physics review of a neutron diagnostic suite for a SPARC-like tokamak [Greenwald \emph{et al} 2018 \href{https://doi.org/10.7910/DVN/OYYBNU}{doi:10.7910/DVN/OYYBNU}]. A notional neutronics model was constructed using plasma parameters from a conceptual device, called the MQ1 (Mission $Q \geq 1$) tokamak. The suite includes time-resolved micro-fission chamber (MFC) neutron flux monitors, energy-resolved radial and tangential magnetic proton recoil (MPR) neutron spectrometers, and a neutron camera system (radial and off-vertical) for spatially-resolved measurements of neutron emissivity. Geometries of the tokamak, neutron source, and diagnostics were modeled in the Monte Carlo N-Particle transport code MCNP6 to simulate expected signal and background levels of particle fluxes and energy spectra. From these, measurements of fusion power, neutron flux and fluence are feasible by the MFCs, and the number of independent measurements required for 95\% confidence of a fusion gain $Q \geq 1$ is assessed. The MPR spectrometer is found to consistently overpredict the ion temperature and also have a 1000$\times$ improved detection of alpha knock-on neutrons compared to previous experiments. The deuterium-tritium fuel density ratio, however, is measurable in this setup only for trace levels of tritium, with an upper limit of $n_T/n_D \approx 6\%$, motivating further diagnostic exploration. Finally, modeling suggests that in order to adequately measure the self-heating profile, the neutron camera system will require energy and pulse-shape discrimination to suppress otherwise overwhelming fluxes of low energy neutrons and gamma radiation.}\\


\end{abstract}

\maketitle

%%%%%%%%%%%%%%%%%%%%%%
%%%%%%%%%%%%%%%%%%%%%%
% Introduction 
%%%%%%%%%%%%%%%%%%%%%%
%%%%%%%%%%%%%%%%%%%%%%

\section{Introduction}

Developments in high temperature superconducting (HTS) magnet technology have opened a pathway toward high-field, compact (HFC) fusion devices. While some notional designs for HTS, HFC tokamaks---like the ARC (Affordable, Robust, Compact) concept \cite{Sorbom2015,kuang2018} and recent SPARC initiative at the MIT Plasma Science and Fusion Center \cite{greenwald2018,mumgaard2018aps,greenwald2018aps,whyte2018aps,lin2018aps,marmar2018aps,white2018aps}---have been put forward, the diagnostic requirements for and operation on such devices have been relatively unexplored. The aim of this paper is to assess the feasibility of neutron measurements on a conceptual SPARC-like device ($R_0 \approx$~1.65~m, $a \approx$~0.5~m) which utilizes HTS magnets to create a magnetic field on axis of $B_0$~=~12~T. For the purposes of this study and that in \cite{creely2018}, this device will be called the MQ1 (\red{Mission $Q \geq 1$}) tokamak. Both opportunities and challenges for plasma diagnosis are found in this HFC parameter space. For instance, high plasma densities and temperatures produce high fusion neutron rates and fluxes; this is optimal for neutron detection, but can also increase nuclear heating, damage, and activation. Moreover, compact size helps lower construction costs of the device, but also limits port space and may even increase the local neutron scattering environment. These are important considerations for diagnostics, especially when some of those explored in this study are similar in size \red{compared} to the tokamak.

A HFC design favors a minimal diagnostic set that is proven, robust, and redundant. \red{Note that the aim of this study is \emph{not} an exhaustive review of all neutron diagnostics or a global optimization of them. Instead,} this paper reviews three widely-used and well-studied neutron diagnostics that support the MQ1 mission, which can be divided into three main objectives: The primary goal is to achieve more power output from fusion reactions ($P_{fus}$) than power input to heat the plasma ($P_{ext}$), thus attaining a minimum fusion gain $Q = P_{fus}/P_{ext}$ of 1. Second, a key technological mission of MQ1 is to demonstrate the operation of HTS magnets in a fusion device. Finally, the physics mission of MQ1 is to explore the self-heating of a plasma with a significant population of fusion alpha particles. For the estimated highest gain scenario on MQ1 of $Q = 3.6$ \cite{mumgaard2016}, approximately 42\% of the total heating power would come from fusion reactions. Therefore, MQ1 provides an opportunity to study near-burning plasma physics, where a ``burning" plasma is defined as 50\% alpha-heated.  Table~\ref{tab:parameters} includes additional machine parameters as well as plasma parameters for a high performance scenario.

\begin{table}[h]
\centering
    %\captionsetup{justification=centering}
    \caption{Machine parameters \cite{greenwald2018,mumgaard2018aps,greenwald2018aps,whyte2018aps,lin2018aps,marmar2018aps,white2018aps,mumgaard2016} of MQ1 and plasma parameters calculated using TSC \cite{creely2018,jardin1986} for a high gain scenario.}
    \label{tab:parameters}
    %\resizebox{\columnwidth}{!}{
    \begin{tabular}{ c  c  c }
         \hline
         Parameter & Symbol & Value \\
         \hline
         Fusion gain & Q & $\sim$3.3 \\
         Aspect ratio & A = R$_0$/a & $\sim$3 \\
         Elongation & $\kappa$ & 1.8 \\ 
         Line-averaged ion density & $\overline{n}_i$ & $\mathrm{3.2 \times 10^{20}\; m^{-3}}$ \\  
         Line-averaged ion temperature & $\overline{T}_i $ & 12.5~keV \\ 
         External heating power & $P_{ext}$ & 30 MW \\
         Fusion power & $P_{fus}$ & 100~MW \\
         Total neutron rate & S & $\sim$3.4$\times 10^{19}$~n/s \\
         Flattop pulse length & $\mathrm{\Delta t_{flattop}}$ & 10~s \\ 
         \hline
    \end{tabular}
    %}
\end{table}

Like other tokamaks, MQ1 would operate in two stages: a deuterium-only plasma phase, followed by a deuterium-tritium (DT) phase. The former phase is discussed in \ref{sec:DD}; the latter phase is the focus of this paper. In these DT plasmas, the neutron-producing fusion reactions of interest are the DD and DT reactions. Other reactions\red{---}like TT and secondary D$^3$He\red{---}will also occur, but at lower rates than DD \cite{Bosch1992}, and will be neglected. For this reason, this study will only consider the DD and DT reactions: 

\begin{equation*}
    \label{DDandDTreactions}
    \begin{split}
            \mathrm{D} + \mathrm{D} \rightarrow \mathrm{n} \, (2.45 \; \mathrm{MeV}) \; + & \; ^3\mathrm{He} \, (0.82 \; \mathrm{MeV}) \\
            \mathrm{D} + \mathrm{T} \rightarrow \mathrm{n} \, (14.1 \; \mathrm{MeV}) \; + & \; \alpha \, (3.5 \; \mathrm{MeV})
    \end{split}
\end{equation*}

In both phases of operation, neutron diagnostics will be vital for measuring the total rate of neutrons produced by MQ1, from which the fusion power can be inferred and fusion gain calculated. In addition, neutron diagnostics can provide key physics insights that bridge the gap between current machines and future fusion power plants, like the ARC design \cite{Sorbom2015,kuang2018}. For example, spatially-resolved neutron measurements can illuminate the alpha-particle birth profile and fusion power density. Additionally, neutron energy spectra can reveal information about the ratio of deuterium to tritium ion density and ion temperature. These measurements are important for tokamak operation as well as the assessment of neutron damage to surrounding machine components, activation of MQ1 infrastructure, and safety of personnel. This paper proposes a neutron diagnostic suite for MQ1 and explores the feasibility of use on the device. The suite incorporates three sets of diagnostics including neutron flux monitors, neutron spectrometers, and neutron camera system. The measurements from these diagnostics analyzed in this study are summarized in Table~\ref{tab:allMeasurements}.

\begin{table*}[h]
\centering
    %\captionsetup{justification=centering}
    \caption{Neutron measurements explored in this study corresponding to each diagnostic and MQ1 mission. Redundant measurements are given in parentheses.}
    \label{tab:allMeasurements}
    \resizebox{\columnwidth}{!}{
    \begin{tabular}{|c|c|c|c|c|c|c|c|}
        \hline
        \multirow{2}{*}{Diagnostic}& \multicolumn{3}{c|}{Mission} \\
         \cline{2-4} & \multicolumn{1}{c|}{Primary Mission } & \multicolumn{1}{c|}{Technology Mission} & \multicolumn{1}{c|}{Physics Mission}\\
         \hline 
         \multirow{2}{*}{Neutron Flux Monitor} & Neutron Rate & \multirow{2}{*}{Neutron Flux/Fluence} &  \multirow{2}{*}{Spatial Asymmetry} \\
          & Fusion Power/Gain & & \\
          \hline
         \multirow{3}{*}{Neutron Spectrometer} & Ion Temperature & \multirow{3}{*}{Energy Spectrum} & \multirow{3}{*}{Alpha Knock-On} \\
          & Fuel Ratio & & \\
          & (Neutron Rate) & & \\
          \hline
         Neutron Camera & (Neutron Rate) & Spatial Neutron Emission & Neutron/Alpha Profile  \\
        \hline
    \end{tabular}
    }
\end{table*}
%\end{center}
%\end{widetext}

The organization of the rest of the paper is as follows: Section~\ref{sec:mcnp} details the neutronics simulations performed for a notional MQ1 tokamak geometry and realistic fusion neutron source. In section~\ref{sec:diagnostics}, the diagnostic suite is presented, with measurements from neutron flux monitors, neutron spectrometers, and neutron cameras explored in subsections~\ref{sec:nfm}, \ref{sec:spec}, and \ref{sec:cam}, respectively. A summary is given in section~\ref{sec:summary}, and an overview of these diagnostics' operation during the MQ1 deuterium plasma phase is discussed in \ref{sec:DD}.

%%%%%%%%%%%%%%%%%%%%%%%%%%%%%%%%%%%%
%%%%%%%%%%%%%%%%%%%%%%%%%%%%%%%%%%%%
% MCNP Modeling
%%%%%%%%%%%%%%%%%%%%%%%%%%%%%%%%%%%%
%%%%%%%%%%%%%%%%%%%%%%%%%%%%%%%%%%%%

\section{MCNP6 modeling of the neutron diagnostics suite}
\label{sec:mcnp}

\begin{figure}[htb!]
    \centering
    \includegraphics[width=0.65\textwidth]{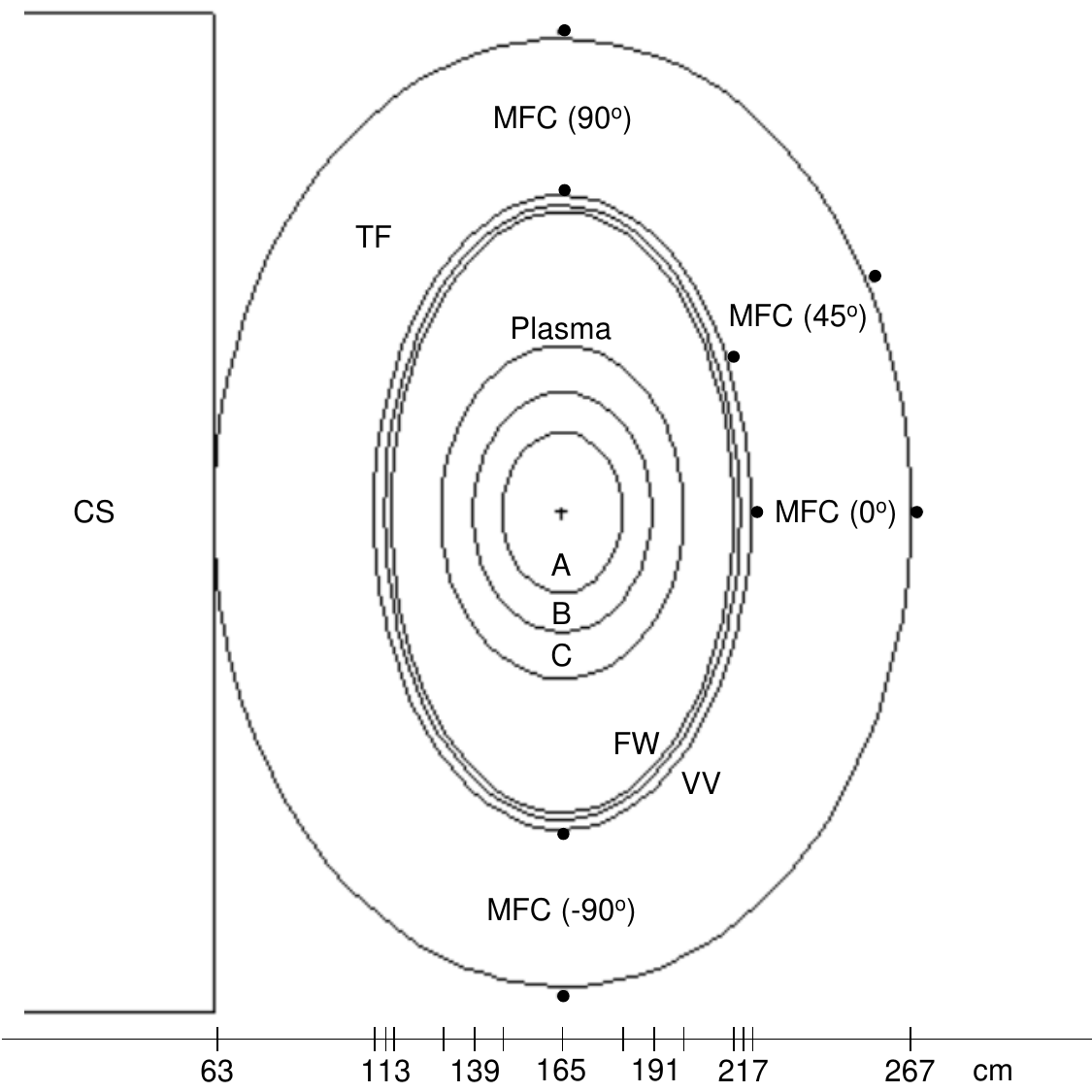}%{figures/poloidalCrossSection_190209.pdf}
    \caption{\red{A poloidal cross-section of the MCNP6 model including the central solenoid (CS), toroidal field (TF) coil, outer vacuum vessel (VV), first wall (FW), and three volumes of the particle source (A, B, and C) as described in Table~\ref{tab:source}. Eight micro-fission chambers (MFCs) are shown as black dots (approximately to-scale) located just outside the VV and TF coil at four poloidal locations: $\theta = 0$, 45, and $\pm90$ degrees. Radial distances from the tokamak center are given in cm at the bottom, with material and geometric details described in the text.}}
    \label{polCrossSection}
\end{figure}

To determine expected signal and background levels for each diagnostic in the neutronics suite, knowledge of neutron transport in a model MQ1 geometry is required. In-situ calibration of diagnostics using well-characterized DD and DT neutron sources is the best mechanism for understanding the tokamak scattering environment; however, as a first estimate, it can be simulated using Monte Carlo modeling techniques. The Monte Carlo N-Particle transport code MCNP6 \cite{goorley2012}, developed by Los Alamos National Laboratory, was used for this task. A notional DIII-D/ASDEX Upgrade-sized geometry was modeled using MQ1 parameters and typical tokamak materials. Figure~\ref{polCrossSection} shows a poloidal cross-section of the MCNP6 model. \red{A midplane cross-section and side view with modeled diagnostics are shown in figures~\ref{specMCNP} and \ref{MCNP_camerasys}.} The vacuum vessel is comprised of two nested elliptical tori: the inner first wall (assumed to be 2~cm of tungsten) and outer support structure (assumed to be 3~cm of Inconel 718). A central solenoid (126~cm in diameter) and 18 toroidal field (TF) coils (47~cm in \red{radial thickness and 22~cm in ``toroidal''} width) were also included in the model, with the TF coils placed toroidally every 20 degrees. Both magnet compositions were simulated as Inconel 718. The ENDF/B-VII cross-section library was used for all nuclear interactions modeled in MCNP6. 

\begin{table*}[h]
\centering
    %\captionsetup{justification=centering}
    \caption{Parameters for the neutron source modeled in MCNP6. Volumes A, B, and C refer to those labeled in figure~\ref{polCrossSection}. The major radius of each torus is $R_0 = 1.65$ m. }
    \label{tab:source}
    \resizebox{\columnwidth}{!}{
    %\begin{tabular}{| c | c | c | c | c | c |}
    \begin{tabular}{ c  c  c  c  c  c }
         \hline
          & Semi-minor & Semi-major & Total fraction & Particle fraction & Temperature \\
        Volume & radius (cm) & radius (cm) & of n/s enclosed & modeled  & modeled (keV) \\
         \hline
         A & 17.5 & 24 & 0.4 & 4/9 & 20 \\
         B & 26.0 & 36 & 0.7 & 3/9 & 15 \\
         C & 35.5 & 50 & 0.9 & 2/9 & 10 \\
         \hline
    \end{tabular}
    }
\end{table*}

%\begin{widetext}
%\centering
\begin{figure*}[h]
    \centering
    \includegraphics[scale=0.45]{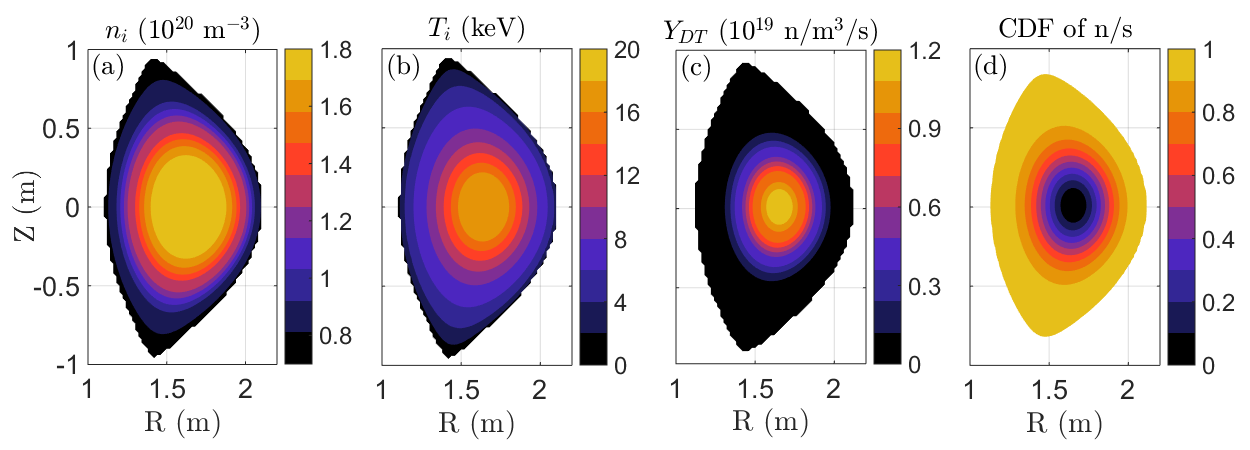}%{figures/ni_Ti_YDT_CDF_180118.png}
    \caption{Profiles of ion (a) density and (b) temperature for the MQ1 high gain scenario used in this study. (c) The profile of DT fusion neutron yield calculated from the profiles in (a) and (b). (d) The cumulative distribution function (CDF) of the fraction of neutrons/s emitted within each surface of constant neutron yield from (c).}
    \label{fig:profiles}
\end{figure*}
%\end{widetext}

A realistic fusion neutron source was also modeled in MCNP6. To produce the source profile, ion density and temperature profiles---as shown in figures~\ref{fig:profiles}(a) and \ref{fig:profiles}(b)---were simulated\footnote{See \cite{creely2018} for details of TSC modeling of MQ1.} using the Tokamak Simulation Code \cite{jardin1986}, with bulk plasma parameters matching those in Table \ref{tab:parameters}. Convolving these profiles with the DT thermal reactivity \cite{Bosch1992} gives the neutron volumetric yield in the form  $Y_{DT} = n_D n_T \langle \sigma v \rangle_{DT}$ (n/m$^3$/s), shown in figure~\ref{fig:profiles}(c). As is seen, the neutron emissivity is highly peaked at the center of the plasma where temperature and density are highest. However, the total neutron emission rate will scale with plasma volume. For this high gain ($Q \approx 3.3$) MQ1 scenario, integrating over the entire plasma volume calculates a total neutron rate of $\sim$3.4$\times 10^{19}$~n/s, approximately one-fifteenth of the highest expected ITER neutron rate \cite{Krasilnikov2005}. In order to most accurately represent the particle source in MCNP6, the plasma was modeled as three concentric tori with parameters given in Table~\ref{tab:source}. The relative fractions of neutrons emitted within each volume was determined using the cumulative distribution function of neutrons generated within the toroidal volume enclosed by a surface of constant yield, as shown in figure~\ref{fig:profiles}(d). The approximately-elliptical surfaces containing 40\%, 70\%, and 90\% were chosen for the MCNP6 particle source and are shown in figure~\ref{polCrossSection}; the neutrons within each volume were given a fusion neutron energy distribution (DD or DT) with ion temperatures of 20, 15, and 10~keV, respectively.

%%%%%%%%%%%%%%%%%%%%%%%%%%%%%%%%%%%%%%%
%%%%%%%%%%%%%%%%%%%%%%%%%%%%%%%%%%%%%%%
% Diagnostics
%%%%%%%%%%%%%%%%%%%%%%%%%%%%%%%%%%%%%%%
%%%%%%%%%%%%%%%%%%%%%%%%%%%%%%%%%%%%%%%
\section{Diagnostic suite}
\label{sec:diagnostics}
%%%%%%%%%%%%%%%%%%%%%%%%%%%%%%%%%%%%%%%%%%%%%%%%%%%%%%%%%%%%%%
% NFM  
%%%%%%%%%%%%%%%%%%%%%%%%%%%%%%%%%%%%%%%%%%%%%%%%%%%%%%%%%%%%%%
\subsection{Neutron flux monitors}
\label{sec:nfm}

Absolutely-calibrated neutron flux monitors (NFMs) provide a direct, passive measurement of the total number of neutrons emitted by the plasma, which can be used to determine the total fusion power and fusion gain $Q = P_{fus}/P_{ext}$, assuming the external heating power is known. An array of NFMs\red{---}typically small in size and having little impact on other diagnostics---can give spatial information of neutron flux and fluence. This data is important for inference of the nuclear damage and heating of the vessel, magnets, diagnostics, and other components. 

The NFM---also known as a neutron yield monitor, neutron detector, or neutron counter---counts the number of neutrons impacting a detector at some location in a given time interval (n/s). For small NFMs, the collection area measures a local neutron flux (n/m$^2$/s), which integrated over the collection time gives neutron fluence (n/m$^2$). A NFM (as well as a spectrometer or camera) will only count a small fraction of the total neutrons emitted by the plasma, so a precise calibration is required to convert neutrons detected to neutrons produced. A DT plasma will primarily generate 14.1~MeV neutrons; however, other sources of neutrons exist: 2.45~MeV DD fusion neutrons, as well as neutrons and photo-neutrons produced when unconfined alphas \cite{Jarvis1994} and gamma radiation from runaway electrons \cite{Gill2000} impact the vessel wall. In this paper, we consider transient-free plasmas with well-confined alphas and therefore focus on neutrons produced by DT and DD reactions.

NFMs have been implemented on many tokamaks including Alcator C-Mod \cite{Fiore1995}, DIII-D \cite{Heidbrink1997}, TFTR \cite{England1986, Hendel1990, Barnes1990, Ramsey1995, Krasilnikov1997}, and JET \cite{JETTEAM1992, Giacomelli2005, Popovichev2004,syme2014,batistoni2017}, among others, and are planned to be used on ITER \cite{Krasilnikov2005, Bertalot2005, Bertalot2012}. Examples of these detectors include BF$_3$ and $^3$He proportional counters \cite{Jarvis1994}, $^{235}$U and $^{238}$U fission chambers \cite{Jarvis1994}, scintillators \cite{Heidbrink1997}, and diamond detectors \cite{Bertalot2005}. In the past, neutron count rates have been digitized at 10-10$^3$~Hz for total neutron production rates of $10^{12}$-$10^{19}$~n/s. On ITER, NFMs are required to measure neutron rates from $10^{14}$-$5\times10^{20}$~n/s with a time resolution of 1 ms \cite{Krasilnikov2005}. The highest performing MQ1 discharge is estimated to produce on the order of $10^{19}$~n/s.

\red{The fission chamber (FC) is one of the most common NFMs used for tokamaks. Its location and size is often determined by the shielding and neutron moderating material required to increase the sensitivity to high energy neutrons. In this study, the \emph{micro-}fission chamber (MFC) \emph{without} a neutron moderator---similar to that planned for ITER---is assessed, primarily motivated by its compact size. Note that future work should explore the use of more general FCs as well as moderating materials.} A detailed overview of MFCs can be found in \cite{Nishitani1998}. A MFC is a pencil-sized, conducting chamber containing an ionizing gas and coated by a fissile\red{/fertile} material, usually $^{235}$U or $^{238}$U. Nuclear reactions of incoming neutrons and the fissile material create $\sim$100~MeV fission fragments which deposit $\sim$40~MeV in the gas, ionizing it and completing the connection between cathode and anode. Gamma rays, however, with energies of $\sim$10~MeV (the same order as the incident neutrons) can also ionize the gas. Therefore, most MFCs are run in Campbelling mode, which utilizes the root-mean-square voltage to discriminate between neutrons and gammas. The maximum MFC reaction rate reported by \cite{Nishitani1998} is $10^{10}$~s$^{-1}$ in Campbelling mode. The approximate detector efficiency (i.e. the fraction of neutrons detected of the total incident) is on the order of $10^{-3}$-$10^{-6}$, depending on the detector. For NFMs, a counting rate as high as possible without saturation is of interest.

In MCNP6, the MFCs were modeled as four concentric cylinders, each of length 7.6~cm, based on specifications from \cite{Nishitani1998}: The innermost ionization chamber, with radius 3.5~mm, is filled with a 95\% argon and 5\% nitrogen ``gas" (i.e. a low density material). The surrounding conducting copper shell is 3.5~mm thick. A 0.3~$\mu$m thick layer of uranium \red{di}oxide (\red{either} $^{235}$UO$_2$ \red{or $^{238}$UO$_2$})\footnote{\red{MCNP6 simulations were performed with both $\Uttf$ and $\Utte$; results were the same within uncertainties.}}, with a total mass of 11 mg, coats the conductor; all are contained within a 4~mm thick aluminum sheath. In total, eight MFCs were modeled at four poloidal locations (directly above and below the plasma, on the outer midplane, and at a poloidal angle of $\sim$45 degrees above the midplane) and two minor radial locations (directly outside the vacuum vessel and directly outside the TF coil), as shown in figure~\ref{polCrossSection}.

Since it is ultimately the interaction of neutrons with the uranium oxide coating of the MFC that leads to neutron counts, the average neutron flux was tallied at this surface for each of the eight MFCs modeled in MCNP6. Assuming that neutrons were primarily coming from the plasma (i.e. few back-scattered neutrons), the neutron rate at the detector was calculated assuming a cross-sectional area of 10.64~cm$^2$ and total fusion neutron rate of 10$^{19}$~n/s. These data are included in Table~\ref{tab:nfmData}. The fluxes inside and outside the TF coils differ by a factor of $\sim$10, and measurements are symmetric (within errors) above and below the plasma, as expected. With detector efficiencies usually in the range of $10^{-3}$-$10^{-6}$ and maximum count rates of $10^{10}$~s$^{-1}$ for Campbelling mode in MFCs \cite{Nishitani1998}, these neutron flux measurements on MQ1 are feasible. In addition, assuming a time resolution of 1 ms, detector efficiency of $10^{-6}$, and NFM measured count rate of $\sim$10$^{13}$~s$^{-1}$, the integrated neutron counts over each time interval is $10^4$ counts, with a statistical uncertainty of only 1\%. Thus, the error will be dominated by calibration which is $\sim$10-20\% in most devices and is discussed below.

\begin{table}[h]
    \centering
    %\captionsetup{justification=centering}
    \caption{MCNP6 neutron rates and statistical errors (in $10^{13}$~n/s) at four poloidal locations (where $\theta=0$ on the outer midplane) for three DT plasma source positions: (i) centered at $R_0$~=~165~cm, $Z$~=~0 and displaced (i) radially by +14~cm or (iii) vertically by +30~cm. Data from measurements outside the TF coil are in parentheses while all others are measurements inside the TF coil. These values assume a total neutron rate of $10^{19}$~n/s emitted by the plasma ($Q \approx 1$) and a 10.64~cm$^2$ cross-sectional area of the uranium oxide coating.}
    \label{tab:nfmData}
    %\resizebox{\columnwidth}{!}{
    \begin{tabular}{| c | c | c | c |}
        \hline
        \multicolumn{4}{|c|}{Neutron Rate ($10^{13}$~n/s)} \\
        \hline
        Poloidal & \multicolumn{3}{c|}{Source position} \\
        \cline{2-4}
        position  & Centered & Radial +14~cm & Vertical +30~cm\\
        \hline
        Below plasma  & 78.7 $\pm$ 0.5 & \multirow{2}{*}{78.2 $\pm$ 0.5} & \multirow{2}{*}{69.7 $\pm$ 0.5} \\
        $\theta = -90^{\circ}$ & (5.5 $\pm$ 0.1) & & \\
        % & [37.5 $\pm$ 0.3] & & \\
         \hline
        Midplane & 93.6 $\pm$ 0.6 & \multirow{2}{*}{102.5 $\pm$ 0.5} & \multirow{2}{*}{86.3 $\pm$ 0.5} \\
        $\theta = 0^{\circ}$ & (8.1 $\pm$ 0.2) & & \\
        % & [51.4 $\pm$ 0.4] & & \\
         \hline
        \multirow{2}{*}{$\theta \approx +45^{\circ}$} & 85.1 $\pm$ 0.6 & \multirow{2}{*}{87.1 $\pm$ 0.5} & \multirow{2}{*}{99.4 $\pm$ 0.6} \\
         & (7.3 $\pm$ 0.2) & & \\
        % & [42.5 $\pm$ 0.4] & & \\
         \hline
        Above plasma & 78.4 $\pm$ 0.5 & \multirow{2}{*}{78.6 $\pm$ 0.6} & \multirow{2}{*}{97.8 $\pm$ 0.7} \\
        $\theta = 90^{\circ}$ & (5.8 $\pm$ 0.1) & & \\
        %& [37.7 $\pm$ 0.4] & & \\
        \hline
    \end{tabular}
    %}
\end{table}

\begin{figure}
    \centering
    \includegraphics[width=0.65\textwidth]{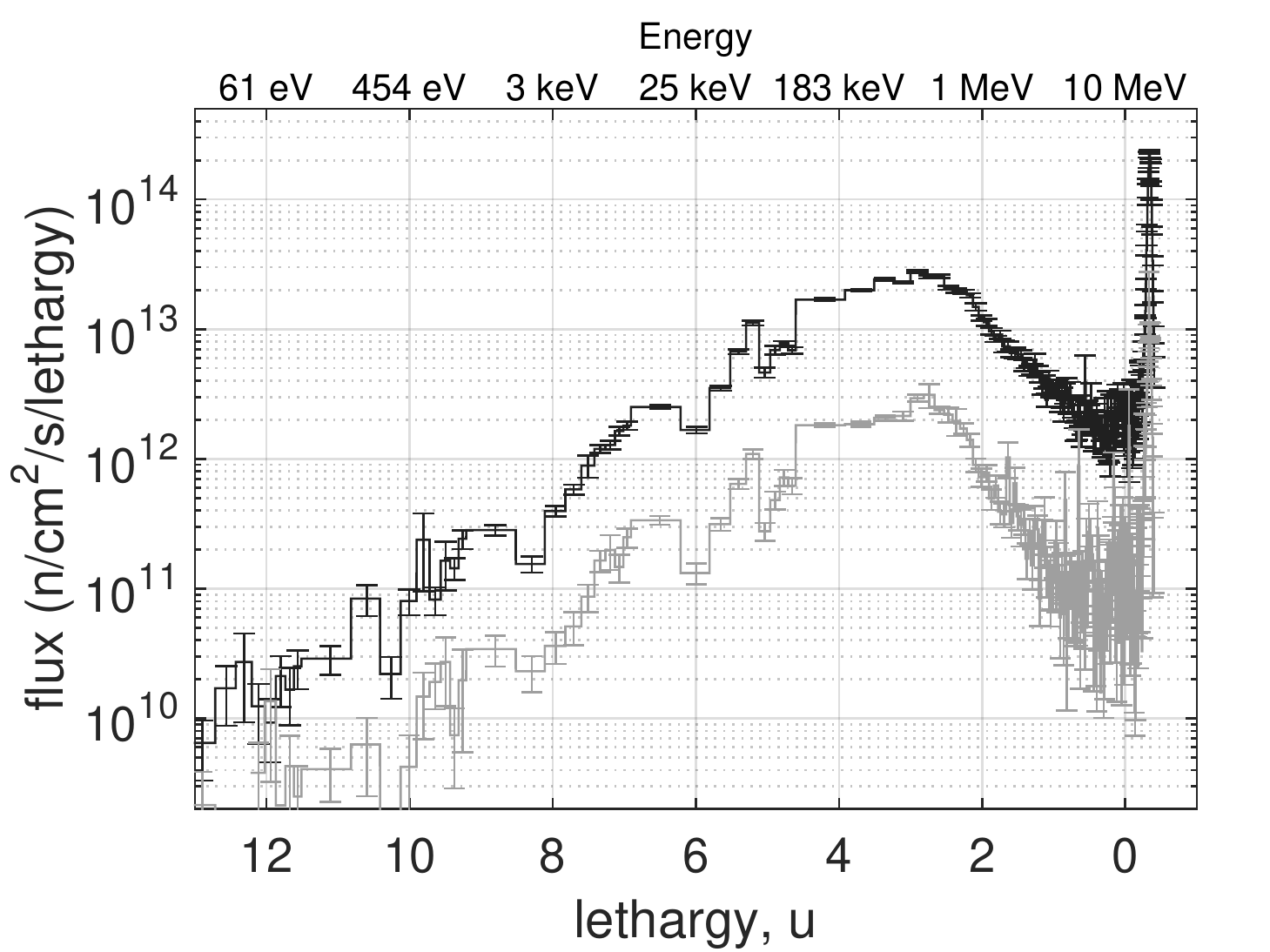}%{figures/MCNP_MFCSpectrum_InnerOuter_Midplane_190217.pdf}
	\caption{Neutron energy spectra, tallied in MCNP6 for DT neutrons only, at the micro-fission chambers (MFCs) located on the midplane ($\theta = 0\deg$) just outside of the vacuum vessel (black) and TF coil (grey). Error bars are given, and a total neutron rate of $10^{19}$~n/s is assumed. The lethargy is u~=~$\ln(E_0/E)$, where $E_0$~=~10~MeV and energies $E$ are labeled at the top. Spectra for the other MFCs are similar, but not shown to reduce clutter.}    
    %\caption{\red{Neutron energy spectra, tallied in MCNP6 for DT neutrons only, at the micro-fission chambers (MFCs) located on the midplane ($\theta = 0\deg$) just outside of the vacuum vessel (black) and TF coil (grey). Error bars are given, and a total neutron rate of $10^{19}$~n/s is assumed. Spectra for the other MFCs are similar, but not shown to reduce clutter.}}
    \label{fig:MFCSpectra}
\end{figure}

\red{The neutron energy spectrum at the MFC surface was also tallied in MCNP6. Two DT neutron spectra for the MFCs located on the midplane outside the vacuum vessel and TF coil are shown in figure~\ref{fig:MFCSpectra}. As expected, the flux decreases by about an order of magnitude for the MFC behind the TF coil compared to that immediately outside the vacuum vessel. As is seen in the figure, a significant fraction of low energy ($<$1~MeV) neutrons impact the MFCs. Because $\Uttf$ has a much higher fission cross-section for thermal neutrons compared to $\Utte$ (which is fissionable by fast neutrons), either $\Utte$ should be used or a neutron moderator should be added to decrease the thermal neutron flux. Future work should investigate the effect of transmutation and decay products on MFC count rates in this high fluence environment.}

The possibility of measuring neutron emission asymmetries using MFCs was also explored. Such asymmetries could result from plasma motion, non-uniform plasma heating, or possibly new physics of near-burning plasmas. To explore the effect of plasma motion, several cases were run in MCNP6 with the plasma at different positions: centered, displaced radially outward by 14~cm (the maximum radial extent), and vertically upward in increments of 10~cm ($\Delta r/a \approx 0.2$). The data for NFMs placed inside the TF coil for the centered and two displaced cases are shown in Table~\ref{tab:nfmData}. Comparing the centered and radially-displaced plasmas, the differences between measurements at the midplane and poloidal angle of 45 degrees are greater than the uncertainty arising from counting statistics, but are less than the expected calibration error. Thus, radial movement is likely undetectable by NFMs. However, for a plasma displaced vertically by 30~cm ($r/a \approx 0.6$), the NFM asymmetry is measurable even when including 10\% uncertainty. Even so, detection of vertical neutron emission asymmetries $<$30~cm are desirable, so the neutron camera is necessary for this capability. Other diagnostics, such as interferometry and magnetics, could make complementary measurements of plasma asymmetries. Note that toroidal asymmetries were not considered in this study.

It is difficult to precisely predict errors in neutron rate measurement for MQ1, but previous devices have detailed them (see Table~\ref{tab:nfmerrors}). Errors can include those due to the calibration sources (usually $^{252}$Cf or DD/DT), conversion between $^{252}$Cf and DD/DT energy spectra, positioning of the calibration source, detector counting statistics, and averaging/fitting of data. The total error is calculated as the sum of the errors in quadrature. Note that the total error in neutron rate measurement for JET was reported to be $\sim$7-10\% \cite{JETTEAM1992,syme2014,batistoni2017}, but no explicit breakdown was given. From these data, we estimate that the uncertainty in total neutron rate will be $\sim$10-15\% in MQ1.

\begin{table}[h]
\centering
    %\captionsetup{justification=centering}
    \caption{Calibration errors reported for several tokamaks.}
    \label{tab:nfmerrors}
    %\resizebox{\columnwidth}{!}{
    %\begin{tabular}{| c | c | c | c |}
    \begin{tabular}{ c  c  c  c }
        \hline
        & Alcator C-Mod \cite{Fiore1995} & DIII-D \cite{Heidbrink1997} & TFTR \cite{Hendel1990} \\
        \hline
        Calibration source rate & 2.6\% & - & 1.5\% \\
        $^{252}$Cf $\rightarrow$ DD/DT calib. & 7.7\% & 12\% & 7\% \\
        Positioning & 2.7\% & $\leq$10\% & 1.5\% \\
        Counting statistics & 6.7\% & 2\% & $\sim$5\% \\
         Averaging/fitting & 14.7\% & - & 9\% \\
        \hline
        Total error & 18.3\% & 15\% & 13\% \\
        \hline 
    \end{tabular}
    %}
\end{table}

The measurements of $Q > 1$ will be of high importance and scrutiny, and the errors in neutron flux measurement will affect the confidence in which the MQ1 team can report the $Q$ attained. Therefore, to estimate the number of independent measurements required to state that the device has achieved its titular goal, a Student's t-test was performed on simulated data. For a given $Q$ value, $N$ data points were randomly selected from a Gaussian distribution with the prescribed mean $Q$ and standard deviation given by the percent error. The average of these data points were then used in a one-sided t-test against the null hypothesis that $Q \leq 1$. If the t-test failed to reject the null hypothesis, $N$ was increased by one, and the process was rerun. If the t-test rejected the null hypothesis with 95\% confidence, $N$ was determined as the number of required measurements of a given $Q$ to determine that the MQ1 mission was achieved. 

Figure~\ref{ttest} shows the results of the t-test as a function of measured $Q$ with two uncertainties: $14\%$, corresponding to a $10\%$ error in the NFM measurement and $10\%$ uncertainty of the external heating power, and 28$\%$, conservatively doubling the error. It is worth noting that this analysis assumes Guassian statistics of the NFM measurements which is not experimentally confirmed, but is a good approximation for large $N$. Furthermore, the analysis is most useful when the device performance is near $Q \approx 1$. If the device exceeds $Q \sim 2$, even with relatively large errors in flux measurement, the team should be confident in the machine achieving its goal. 

\fig{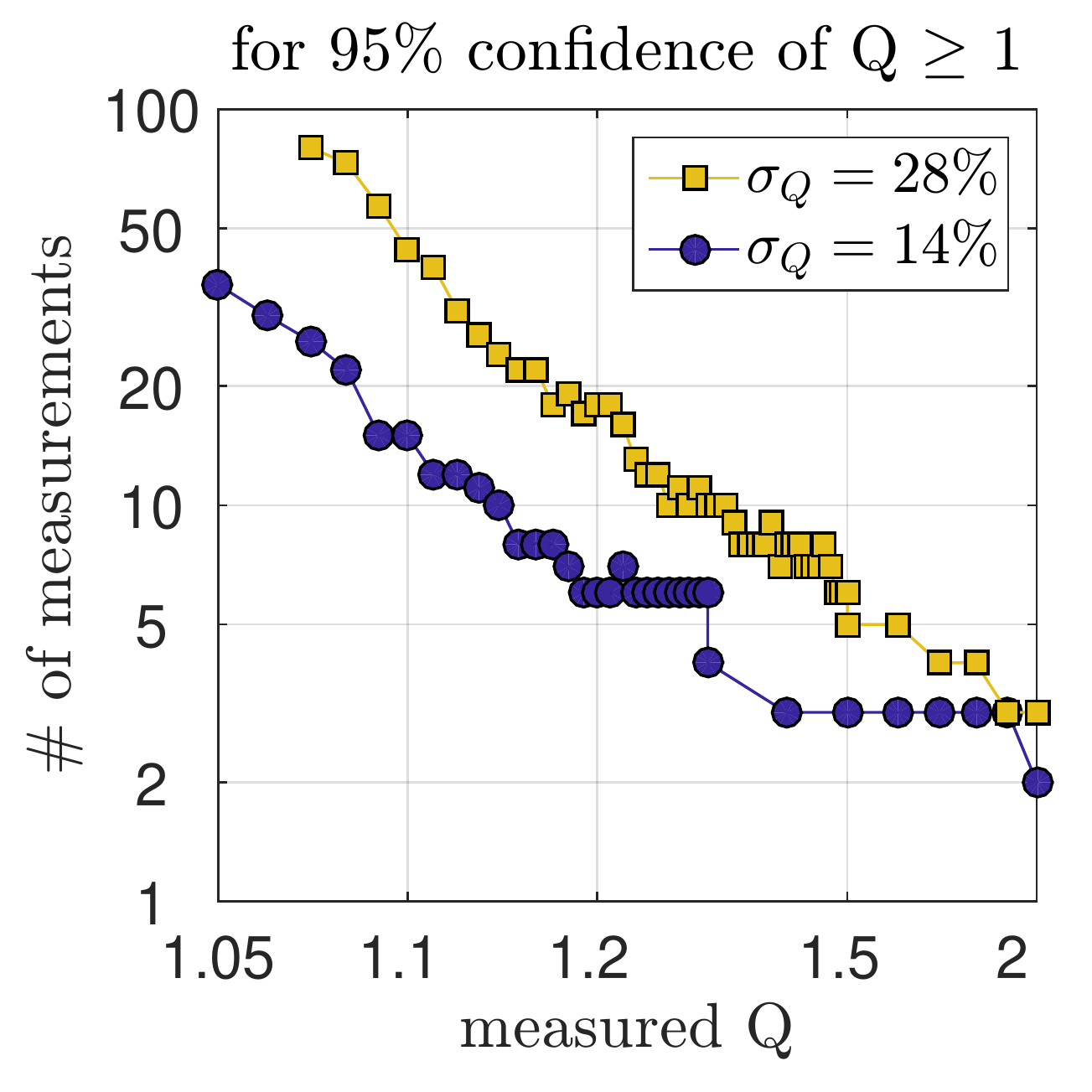}{Number of measurements of a given $Q$-value required to claim 95\% confidence of $Q \geq 1$ for \emph{total} errors of 14\% and 28\%.}{ttest}
%\fig{figures/Alex_ttest.pdf}{Number of measurements of a given $Q$-value required to claim 95\% confidence of $Q \geq 1$ for \emph{total} errors of 14\% and 28\%.}{ttest}

%%%%%%%%%%%%%%%%%%%%%%%%%%%%%%%%%%%%%%%%%%%%%%%%%%%%%%%%%%%%%%
% Spectrometer %%%%%%%%%%%%%%%%%%%%%%%%%%%%%%%%%%%%%%%%%%%%%%%
%%%%%%%%%%%%%%%%%%%%%%%%%%%%%%%%%%%%%%%%%%%%%%%%%%%%%%%%%%%%%%
\subsection{Neutron spectrometers}
\label{sec:spec}

The neutron spectrometer diagnostic provides time and energy-resolved measurements of neutron flux integrated along a line-of-sight through the plasma. Characteristics of the measured energy distribution, such as the Doppler broadening of peaks or high energy tail, can determine the ion temperature and explore fast ion effects. Because DD and DT reactions produce neutrons of different energies, these spectrometers can measure relative densities of deuterium and tritium fuels. In addition, absolute calibration provides a redundant measurement of total neutron production in the plasma.

\begin{figure}[h!]
    \centering
    \includegraphics[width=0.65\textwidth]{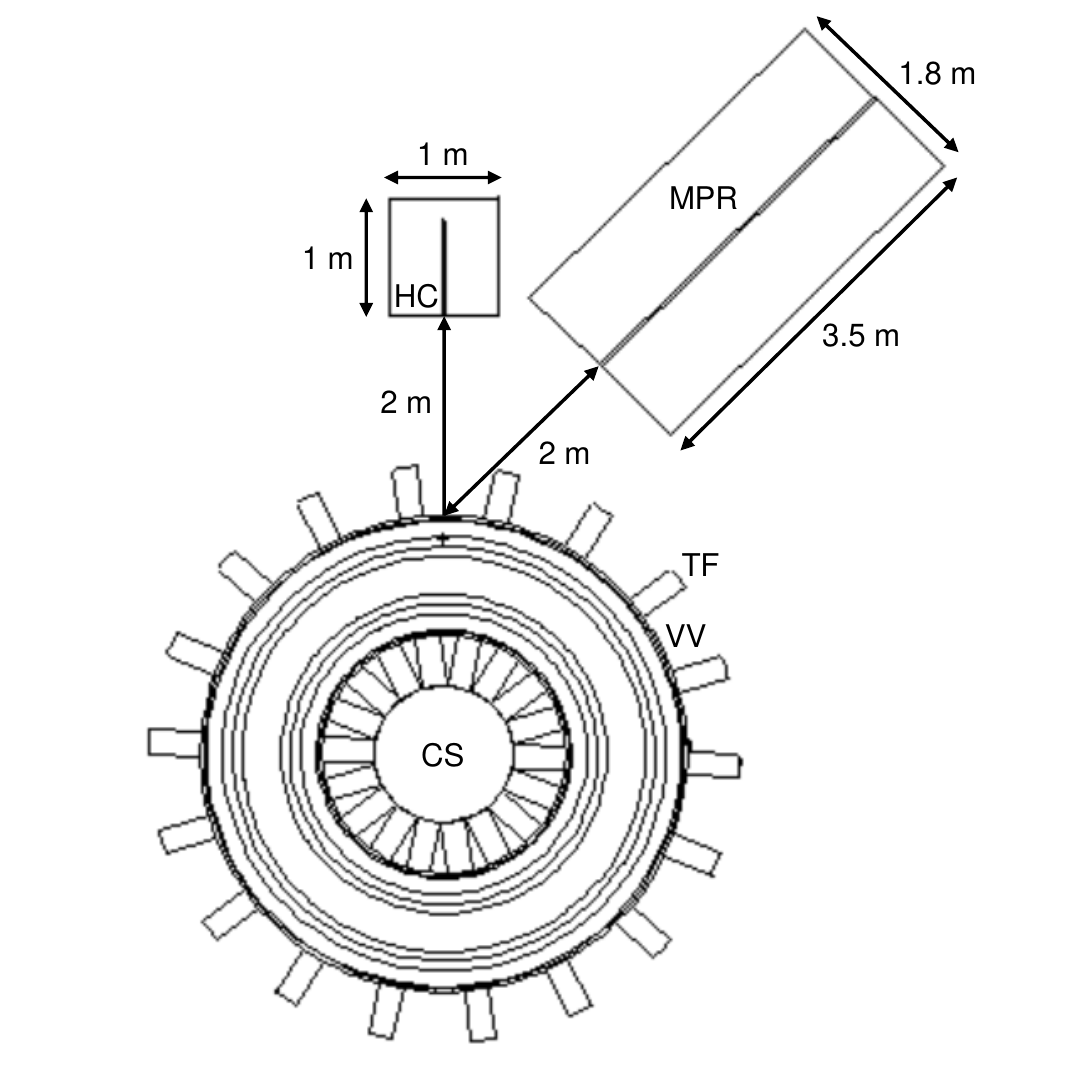}%{figures/topDownTangentialMPR_190209.pdf}
    \caption{\red{A top-down midplane cross-section of the tokamak, tangential MPR spectrometer, and horizontal neutron camera (HC) modeled in MCNP6. Note that the radial MPR is not shown. Also labeled are the central solenoid (CS), outer vacuum vessel (VV), and TF coil. Distances between the apertures and dimensions of diagnostics are indicated, with other material and geometric details provided in the text.}}
    \label{specMCNP}
\end{figure}

Neutron spectrometers in various configurations have been used on TFTR \cite{osakabe,strachan1988neutron} and JET \cite{johnson20082,kallne1999new}, among other devices \cite{sasao}, and are planned for ITER \cite{Murari}\red{; although, the current design for ITER's high resolution spectrometer is left unspecified \cite{Bertalot2012}.} There are multiple approaches to diagnose neutron energy spectra, many of which show promise for near-burning plasma experiments \cite{sunden2013evaluation}; examples are magnetic proton recoil (MPR) \cite{frenje1999}, solid state diamond detection \cite{DiamondAng}, time-of-flight measurements \cite{johnson20082}, and scintillators coupled to pulse-shape discrimination systems \cite{TTC}. In particular, the MPR system has demonstrated impressive performance; the system at JET has operated for nearly two decades and successfully measured all spectroscopic quantities of interest in this study \cite{sunden2013evaluation}. \red{One of the major drawbacks of the MPR system is its relatively large size; here, the MPR is almost as the same size as the tokamak, as seen in figures~\ref{specMCNP} and \ref{MCNP_camerasys}. However, as MQ1 is a conceptual device, space can be allotted for such a large, high priority diagnostic. Furthermore, the philosophy of MQ1 favors robust and proven diagnostics to rapidly achieve the $Q \geq 1$ mission.} Therefore, for MQ1, an MPR spectrometer is \red{assessed} due to its demonstrated high sensitivity and resolution.

An MPR spectrometer consists of a shielded chamber which houses an electromagnet and detector array. The applied magnetic field is perpendicular to a collimated view through which neutrons stream. A fraction of neutrons elastically scatter with a proton conversion foil, ejecting protons of similar energy (within 0.5\% due to the similar mass and collimation \cite{sunden2009thin}). The scattered protons are spatially dispersed in the magnetic field by a distance proportional to their Larmor radius, $r_L = \sqrt{2 m E/eB}$, where $m$ and $e$ are the respective mass and charge of the proton, $E$ is the proton energy, and $B$ is the strength of the MPR's magnetic field. Protons of different energies will impact detectors---scintillators coupled to photomultiplier tubes---at different positions within the spectrometer. This spatial proton profile can be interpreted as a neutron energy spectrum. 

In order to estimate the neutron flux reaching the conversion foil as well as background/noise at the detector, two MPR-like spectrometers were modeled in MCNP6 (see figure~\ref{specMCNP}) using specifications from \cite{sjostrand2006new}. For each, the magnet/detector chamber is modeled as an empty box of height 1.5~m, length 1.5~m, and width 0.8~m, shielded on all sides by concrete that is 0.5~m thick, except for the face nearest the plasma which has thickness 1.5~m. At the top of the magnet chamber is a square collimator of cross-sectional area 10~cm$^2$ which extends through the entire length (3.5~m) of the spectrometer. One spectrometer is situated such that its collimated line-of-sight is tangential to the magnetic axis on the midplane, thereby maximizing the total number of neutrons viewed \red{(see figure~\ref{specMCNP})}; the other has a midplane radial view of the plasma \red{(see figure~\ref{MCNP_camerasys})}. Both apertures for the spectrometers (2~m from the collimator opening) are modeled as voids in the vacuum vessel outer support structure so that the wall consists of only the first wall (2~cm of tungsten) in that location. The aperture for the radial spectrometer is 15~cm $\times$ 15~cm; the tangential aperture has the same width but an increased height of $\sim$64~cm to accommodate the horizontal neutron camera lines-of-sight, as discussed in section~\ref{sec:cam}.

For this setup, a magnetic field strength of $B$~=~0.9~T could be used to confine protons up to 20~MeV in energy.  In addition, the energy resolution of the spectrometer, as a function of the particle energy $E$ and detector spatial resolution $\Delta x$, is $\Delta E \approx e B \Delta x \sqrt{E/2m}$. For detectors spaced $\Delta x$~=~1~cm, the energy resolutions are $\sim$100~keV and $\sim$230~keV at energies of 2.5~MeV and 14.1~MeV, respectively. To examine smaller ranges of the energy spectra, higher resolution can be achieved on the MPR system by adjusting the magnetic field.

\red{As mentioned,} the MPR is a large diagnostic, similar to the size of the tokamak itself. In this setup, there is 2~m of available distance between each spectrometer and its aperture, which should allow for other diagnostics as long as they do not block the MPR line-of-sight. Predicted signal levels, as will be described, indicate that the MPR could be placed farther away (up to $\sim$5~m) and still measure adequate neutron counts. \red{Another consideration is the effect of the MPR magnetic field on that of the tokamak. Assuming that the MPR produces a roughly dipole magnetic field, the field strength will decrease with distance from the MPR system like $r^{-3}$. Thus, for an MPR located 2~m from the tokamak, the stray field at the tokamak is reduced by almost an order of magnitude. Due to the high field of the tokamak itself, this introduces an error field which is a factor of $\sim$100 smaller than the tokamak's field. This 1\% error field is below acceptable limits on the ripple field \cite{goldston1981confinement} and below error field experiments performed on DIII-D \cite{schaffer2011iter}. Therefore, the error field due to the MPR is not a concern at this stage of diagnostic development and could be reduced by moving the diagnostic farther away from the tokamak.} 

From the energy spectrum, the MPR system is capable of measuring a host of parameters, including basic plasma parameters \red{and} effects of \red{radio frequency} heating \cite{nocente2011calculated}. This paper will primarily focus on achieving measurements of ion temperature, fuel ratio, and alpha knock-on neutron emission, each detailed in the following subsections. These three measurements not only support the MQ1 missions, but also illustrate the many measurement capabilities of the MPR. Both ion temperature and fuel ratio are indicative of plasma performance and can influence tokamak operations like external heating and fueling. While ion temperature is measurable in several ways, the fuel ratio can currently only be inferred through neutron diagnostics. Both measurements require a time resolution less than the energy confinement time, which is estimated to be $\geq$350 ms in MQ1 \cite{mumgaard2016}. The desired accuracy for ion temperature and fuel ratio measurements are \red{within} $<$10\% and $<$20\%, respectively. 

The final measurement of interest is the high energy tail of the neutron energy spectrum resulting from the phenomenon of alpha knock-on neutron (AKN) emission. Such a measurement can test classical confinement of hot ions and has only been successfully observed once on JET with limited statistics \cite{kallne2000observation}. The MPR's ability to measure fine spectral details, such as AKN, can provide insight into the new physics and confinement of large populations of fusion alphas in the MQ1 plasma\red{; these are} critical measurements for the physics mission as well as next-step devices.

For the following analyses, several assumptions are made of the MPR system on MQ1, all of which have been demonstrated on JET. First, we assume that the MPR system is capable of a neutron detection efficiency of $10^{-5}$ \cite{sunden2009thin}. Furthermore, we assume a minimum energy resolution of 100~keV, measurable energy range from 1-20~MeV, and maximum allowable neutron count rate at the detector of $10^8$~s$^{-1}$ \cite{sunden2013evaluation}.

%%%%%%%%%%%%%%%%%%%%%%%%%%%%%%%%%%%%%%%%%%%%%%%%%%%
\subsubsection{Ion Temperature}

The ion temperature $T_i$ can be determined by measuring the Doppler broadening of the DT neutron spectrum. The DT peak at 14.1~MeV is used instead of the DD peak at 2.45~MeV because of the $\sim$100$\times$ larger reactivity and likelihood that the energy down-scattered DT neutrons will overwhelm the DD peak. The neutron energy distribution, assuming a Maxwellian of temperature $T_i$, is given by \cite{hutchinson2002}

\begin{equation}
    P(E_n)dE_n \propto \exp( (E_n -\overline{E}_n)^2/2\sigma^2),
\end{equation}

\noindent where $\overline{E}_n$~=~14.1~MeV is the average neutron energy for the DT reaction, and $\sigma$ is the standard deviation given as a function of $T_i$,

\begin{equation}
    \sigma^2 = \frac{2 m_n \overline{E}_n}{m_n+m_\alpha} T_i.
\end{equation}

\noindent Here $m_n$ and $m_\alpha$ are the neutron and alpha particle masses, respectively. To zeroth order, the DT peak can be fit with a Gaussian to determine $T_i$ \cite{ballabio1998}. However, during the use of neutral beam or ion cyclotron heating, the ion population is non-thermal, and the Maxwellian approximation is no longer valid. Still, the thermal population of neutrons is distinguishable in the spectra, allowing measurements of $T_i$ but requiring increased computation time \cite{kallne1999new,tardocchi2004ion}.

Assuming a total DT fusion neutron rate of 10$^{19}$~n/s, MCNP6 results indicate that $\sim$10$^{13}$~n/s would impact the proton-conversion foil, producing $\sim$10$^8$ protons/s at the conversion foil, of which $\sim$10$^6$ protons/s would hit each detector. This approaches the expected performance limits of the MPR. \red{However, in practice the number of counts reaching the detector could be reduced by using a smaller collimator, reducing the foil thickness, or moving the MPR farther from the tokamak, as discussed previously.} In the current setup, the count rate is sufficiently high, allowing a time resolution of 10~ms and total counting error of $\sim$1\%. An additional source of error in calculating $T_i$ comes from the finite energy resolution of the spectrometer. The fractional error in the calculated ion temperature is approximated as 

\begin{equation}
    \frac{\Delta T_i}{T_i} = \sqrt{\frac{2}{N}}\left(1+\frac{T_r}{T_i}\right),
\end{equation}

\noindent where $N$ is the number of counts in the spectra used, and $T_r$ is the apparent thermal Doppler broadening which accounts for the finite resolution of the detector as well as other systemic uncertainties in the device \cite{tardocchi2004ion}. A typical value for the thermal Doppler broadening on the MPR system at JET is $T_r \approx$~11~keV \cite{tardocchi2004ion}. Using this value and 10~ms resolution, the fractional error predicted for MQ1 is $\leq$10\%. Estimating the uncertainty introduced from calibration on MQ1 is difficult; previous systems have demonstrated performance of $\sim$10\% error \cite{sunden2013evaluation}. Thus, calibration error is expected to dominate.

\begin{figure}
    \centering
    \includegraphics[width=0.65\textwidth]{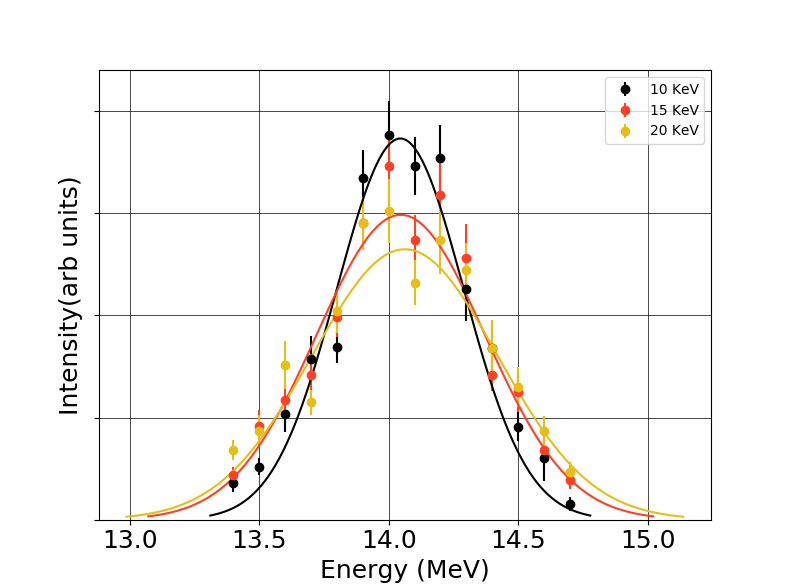}%{figures/RadTiErrorSqrd.png}
    \caption{MCNP6 synthetic spectral data (with error bars), as measured by the radial spectrometer, are shown for three uniform plasma temperatures: $T_i$~=~10 (black), 15 (red), and 20~keV (yellow). Gaussian fits (solid lines) estimate $T_i$, as given in Table \ref{tab:Ti}.}
    \label{fig:temp}
\end{figure}

\begin{table}[]
    \centering
    \caption{Synthetic measurements of $T_i$ (keV) from the radial and tangential spectrometers with the given MCNP6 source profile. Uniform profiles are modeled with $T_i$ constant throughout the entire plasma; the full profile is that described in Table \ref{tab:source}.}
    \label{tab:Ti}
    \begin{tabular}{c c c}
        \hline
        $T_i$ (keV) & \multicolumn{2}{c}{Spectrometer} \\
        source profile & Radial & Tangential  \\
        \hline
        Uniform 10 & 10.6 & 12.3 \\
        Uniform 15 & 18.7 & 18.4 \\
        Uniform 20 & 22.7 & 22.2 \\
        Full & 16.9 & 18.4 \\
        \hline
    \end{tabular}
\end{table}

To demonstrate the feasibility of a $T_i$ measurement on MQ1, four MCNP6 simulations were performed, each with a different temperature profile. The neutron energy spectrum was tallied in the collimator leading to the proton conversion foil with an energy resolution of $\Delta E = 100$~keV from 1-15~MeV (attainable in our MPR setup with a magnetic field of $B \approx 0.4$~T). A least-squares Gaussian fit of the data, weighted by the inverse square of the error, was used to calculate the temperature. Figure~\ref{fig:temp} shows MCNP6 data from the radial spectrometer for plasmas with uniform profiles of temperature: $T_i$~=~10, 15, and 20~keV. The fitted temperatures for both spectrometers are displayed in Table \ref{tab:Ti}. Note that this calculation consistently overpredicts the uniform temperature by $\sim$10-20\%, whereas an \emph{underprediction} of \red{$\sim$}20\% by a single line-of-sight radial spectrometer is reported in \cite{marocco2011}. One probable cause for this is spectral broadening from energy down-scattered neutrons. For the full radial profile (as described in Table \ref{tab:source}), the spectrometers measure a line-averaged temperature; as expected, the estimation is lower for the radial compared to the tangential spectrometer, since the latter views a larger fraction of high $T_i$ regions. Actual experimental neutron spectra will be affected by many parameters---plasma rotation, radial profiles, external heating, etc.---so proper calibrations must be performed, and more advanced models should be used for fitting.

%%%%%%%%%%%%%%%%%%%%%%%%%%%%%%%%%%%%%%%%%%%%%%%%%%%
\subsubsection{Fuel Ion Ratio}

For a measurement of the fuel ion ratio, the number of DD and DT neutrons produced by the plasma must be compared\red{, requiring simultaneous measurements of the spectral peaks at 2.5 and 14.1~MeV, respectively.} This section will \red{consider} only thermal plasmas. Spectra with significant non-thermal populations can be analyzed but require further modeling of the plasma \cite{sunden2013evaluation}. The intensity of the fusion neutron peaks are proportional to the yield

\begin{equation}
    Y_{ij} =\frac{n_i n_j}{1+\delta_{ij}}\langle\sigma v \rangle_{ij},
\end{equation}

\noindent where $i$ and $j$ represent the fusion reactant species, and $\delta_{ij}$ is the Kronecker delta function accounting for double-counting of like-species. Thus, for DT plasmas, the fuel ratio can then be found by taking the ratio of the DT and DD intensities, giving

\begin{equation}
    \frac{n_T}{n_D} = \frac{1}{2} \frac{Y_{DT}}{Y_{DD}}\frac{\langle\sigma v \rangle_{DD}}{\langle\sigma v \rangle_{DT}}.
\end{equation}

When using one spectrometer sensitive to both neutron energies, the challenges are (i) setting the magnetic field strength and detector geometry to measure both energies with sufficient resolution to determine the temperature and hence the reactivities, as well as (ii) distinguishing the DD energy peak above the down-scattered DT spectrum \cite{kallne1997feasibility}. It is conceivable to measure both peaks in a detector of the size described; \red{however, it may require an unacceptable loss in resolution, in which case it may be necessary to include a spectrometer specialized for measuring the DD peak.} The exact design and optimization of the MPR are left for future work. Further discrimination is required of background events, such as stray protons and gamma rays which reach the detectors. MPR systems have demonstrated discrimination of signal from background events by a factor of 10:1 around the DD peak \cite{sunden2009thin}. Thus, the limitation on the fuel ratio measurement is assumed to arise from distinguishing the DD peak from the down-scattered DT neutrons.

\begin{figure}[h!]
    \centering
    \includegraphics[scale=0.65]{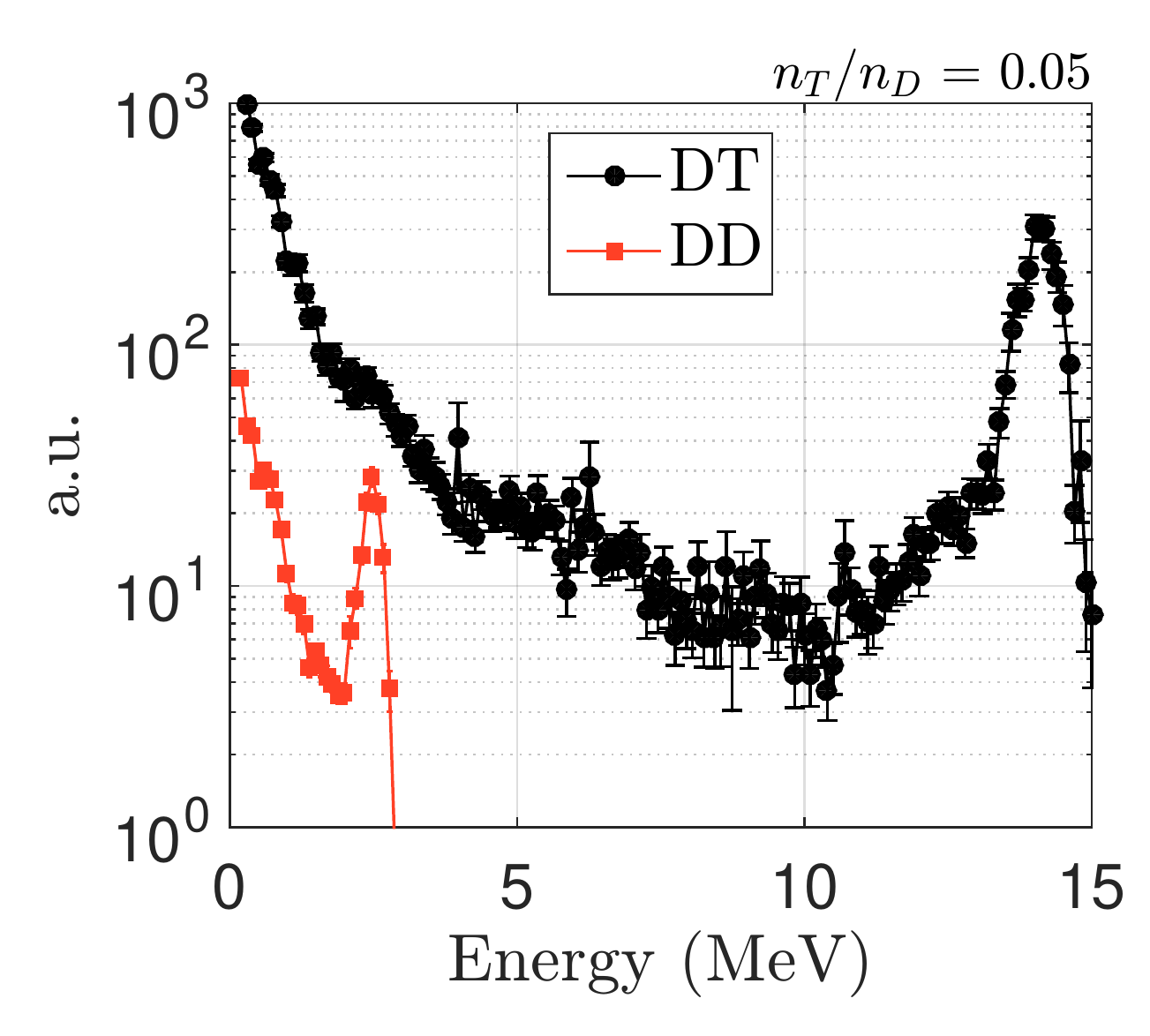}%{figures/neutronEnergySpectra_DD_DT_nTOvernD5e-2_180516.pdf}
    \caption{MCNP6 neutron energy spectra tallied in the neutron spectrometer collimator volume for DD (squares) and DT (circles) neutron sources. Each spectrum has been scaled such that the fuel-ion ratio is $n_T/n_D = 0.05$ for which the DD peak at $\sim$2.5~MeV is 1/3 of the down-scattered DT spectrum (i.e. 1/4 of the summed spectra). MCNP6 data suggests fuel ratios above 0.06 will not be measurable with the current setup.}
    \label{spectra}
\end{figure}

For MQ1, the distinguishability of the DD peak above the down-scatter was determined using the MCNP6 Doppler-broadened data. Figure~\ref{spectra} shows the MCNP6 energy spectra for DT and DD neutrons scaled\footnote{Separate MCNP6 simulations were performed for DD and DT fusion sources, so the results were scaled appropriately for comparison.} to a fuel ratio of $n_T/n_D = 0.05$. The characteristic down-scatter of DT neutrons is seen in the low energy range. The DD peak was assumed to be distinguishable if its contribution to the \emph{total} spectrum (DD + DT) was 25\% at 2.45~MeV. This number is similar to previous reports on the feasibility of fuel ion ratio measurements \cite{kallne1997feasibility}. Using the 25\% value, the \red{upper} limit for calculating the fuel ratio was found to be about $n_T/n_D \approx 6\%$, a value useful in trace tritium experiments. Note that the MCNP6 data used for this calculation was volume-averaged within the MPR collimator due to poor counting statistics at the conversion foil interface. In reality, with a large neutron rate, we expect reduced energy down-scatter in the spectrum of neutrons reaching the foil surface, making this estimate conservative.

Of concern, this result suggests that an MPR will not be able to provide fuel ratio measurements during full DT operation. Synthetic data studies on ITER have suggested that the fuel ratio with an MPR system can operate up to $n_T/n_D = 60\%$ \cite{sunden2013evaluation}, assuming a near radial view of the plasma. A tangential view is expected to provide a higher possible fuel ratio measurement by at least a factor of 3 \cite{kallne1997feasibility,antozzi1996neutron}. However, this is not observed in the MCNP6 simulations performed for MQ1. For the radial neutron camera on ITER, the down-scattered DT spectral counts at $\sim$2.5~MeV are approximately 30 times fewer than the counts at the DT peak \cite{ericsson2010neutron}. In comparison, this study finds only 4 times fewer DT counts at 2.5~MeV compared to 14.1~MeV. In \cite{kallne1997feasibility}, the majority of down-scattered neutrons were concluded to come from reflections off the far wall along the spectrometer line-of-sight. However, removing this part of the wall in our simulations---to approximate a ``neutron absorber"---had no effect\red{, indicating that even a view dump would not help to distinguish the DD peak.} Higher fidelity simulations should be performed to improve upon errors, and other diagnostic techniques should be explored.

%%%%%%%%%%%%%%%%%%%%%%%%%%%%%%%%%%%%%%%%%%%%%%%%%%%
\subsubsection{Alpha Knock-On Neutron Emission}

The AKN effect is one measurement accessible by the MPR which explores the new physics of a near-burning plasma: With low probability, a confined DT fusion alpha particle can collide head-on with either a deuteron or triton, boosting the fuel ion to suprathermal energies of $\sim$3.2 and 3.5~MeV, respectively. This suprathermal ion can then undergo DT fusion with thermal reactants producing a suprathermal neutron with energy $\leq$20~MeV. As the resulting energy spectra depend on alpha confinement, a measurement of the AKN spectra can test the classical assumptions of fast ion confinement \cite{kallne2000observation}. If measured on a relevant timescale, the AKN effect can also be used as a measurement of alpha particle pressure \cite{ballabio1997alpha}.

To determine the feasibility of an AKN measurement in MQ1, we compare AKN production and detection to JET, which measured this effect on a limited number of its highest powered discharges \cite{kallne2000observation}. The proposed MPR spectrometer for MQ1 has the same resolution properties as that on JET; however, the neutron \red{flux} is expected to be a factor of $\sim$100 greater than the highest performing discharges on JET. Furthermore, the fractional contribution of AKN to the energy spectrum increases with temperature, improving MQ1 counting statistics by a further factor of $\sim$100 \cite{ballabio1997alpha}. For a time resolution of 1 s, we expect that an AKN measurement on MQ1 will have improved statistics by a factor of 1000 compared to previous experiments. This time scale allows the study of alpha dynamics over the $\sim$10 second discharge length of MQ1.

%%%%%%%%%%%%%%%%%%%%%%%%%%%%%%%%%%%%%%%%%%%%%%%%%%%%%%%%%%%%%%
% Camera  %%%%%%%%%%%%%%%%%%%%%%%%%%%%%%%%%%%%%%%%%%%%%%%%%%%%
%%%%%%%%%%%%%%%%%%%%%%%%%%%%%%%%%%%%%%%%%%%%%%%%%%%%%%%%%%%%%%
\subsection{Neutron camera system}
\label{sec:cam}

Neutron cameras provide spatially, temporally, and spectrally-resolved measurements of neutron emission. As neutrons are born within the plasma, they stream through a thin aperture and are collimated in steel channels. Each channel terminates with a fast neutron detector that typically consists of a liquid scintillator and photomultiplier tube which provide a time and energy-resolved measurement of neutron emissivity corresponding to the line-of-sight viewed by the channel. Because neutrons and alpha particles are born simultaneously, the alpha birth profile is obtained indirectly from the neutron emission profile \cite{ishikawa2002first}. Thus, a camera system could diagnose the fusion power density profile and resulting self-heating of MQ1's near-burning plasma; both provide key information for future high-field fusion devices such as ARC \cite{Sorbom2015,kuang2018}. In JET, the neutron camera system measured the fuel ratio profile during a series of trace tritium experiments; sawtooth crashes were also observed as temperature and density profile flattening in the core caused drops in DT neutron rates \cite{Marcus1992}. Redundant measurements made by the neutron camera system include the total fusion neutron rate and $T_i$ profile \cite{marocco2011}, given adequate signal levels.

Neutron camera systems have been demonstrated on a variety of different machines: Single camera systems were employed on TFTR \cite{sasao,roquemore1990tftr,hendel1985collimated}, JT-60U \cite{ishikawa2002first}, and MAST \cite{Cecconello2012}; a two camera system has been used on JET \cite{Adams1993}; and a three camera system is being designed for ITER \cite{Bertalot2012}. JET's nine-collimator vertical array and twelve-collimator horizontal array allow for 2D tomographic reconstructions \cite{mll2008} of the plasma and wide imaging coverage of a poloidal cross-section. A similar design is proposed for MQ1. However, to keep open the option of a double-null magnetic geometry, an upper camera that is slightly off-axis could be used, similar to MAST and ITER designs \cite{Bertalot2012,Cecconello2012}. This complements a horizontal camera which sits on the midplane of the tokamak.

\begin{figure}
    \centering
    \includegraphics[scale=0.4]{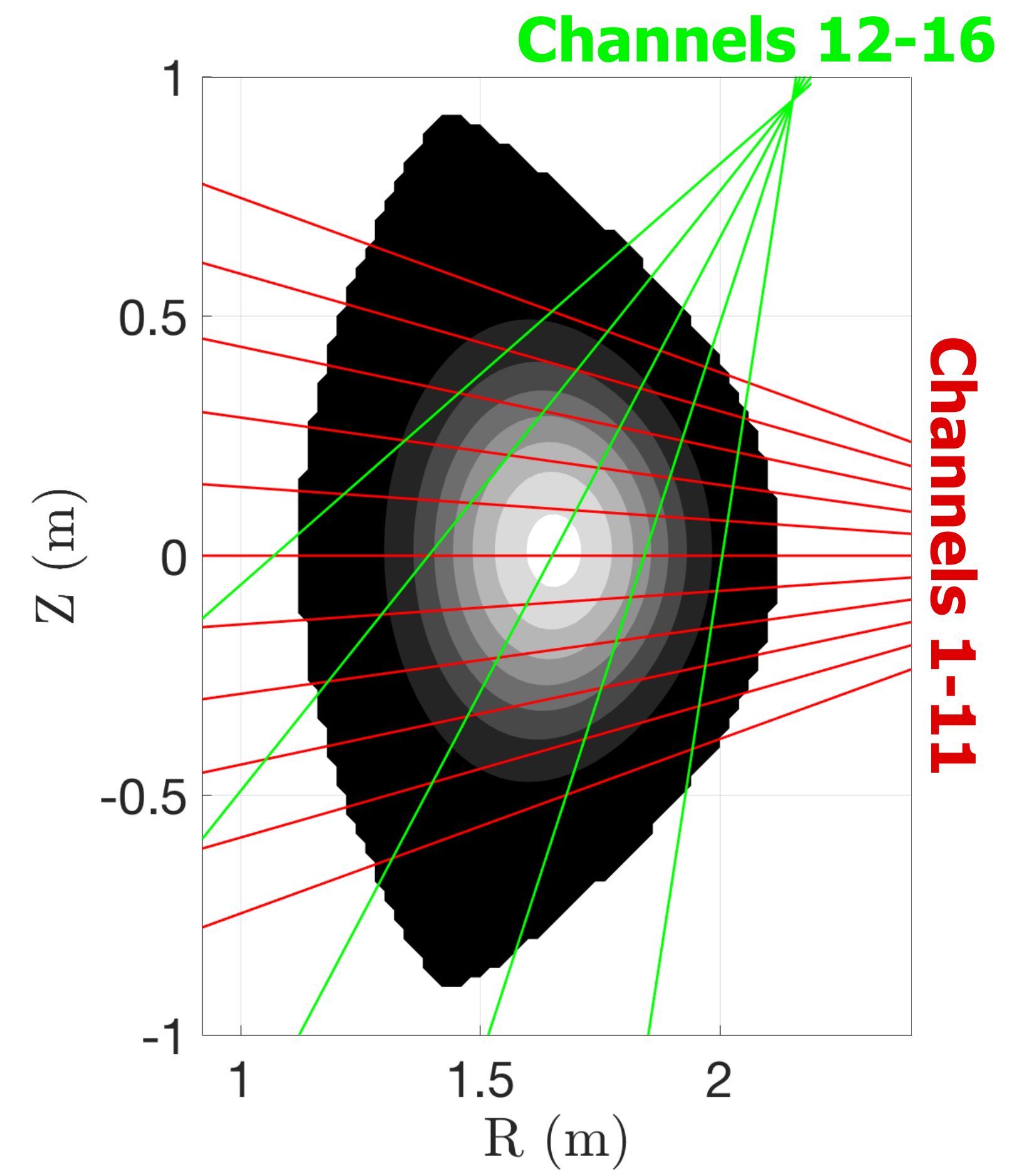}%{figures/neutronCameraLinesOfSight.pdf}
    \caption{Neutron camera lines-of-sight from the horizontal camera (Channels 1-11) and upper camera (Channels 12-16) overlay the DT fusion yield from figure \ref{fig:profiles}(c).}
    \label{syntheticCamera}
\end{figure}

\begin{figure}
\centering 
  \includegraphics[scale=0.65]{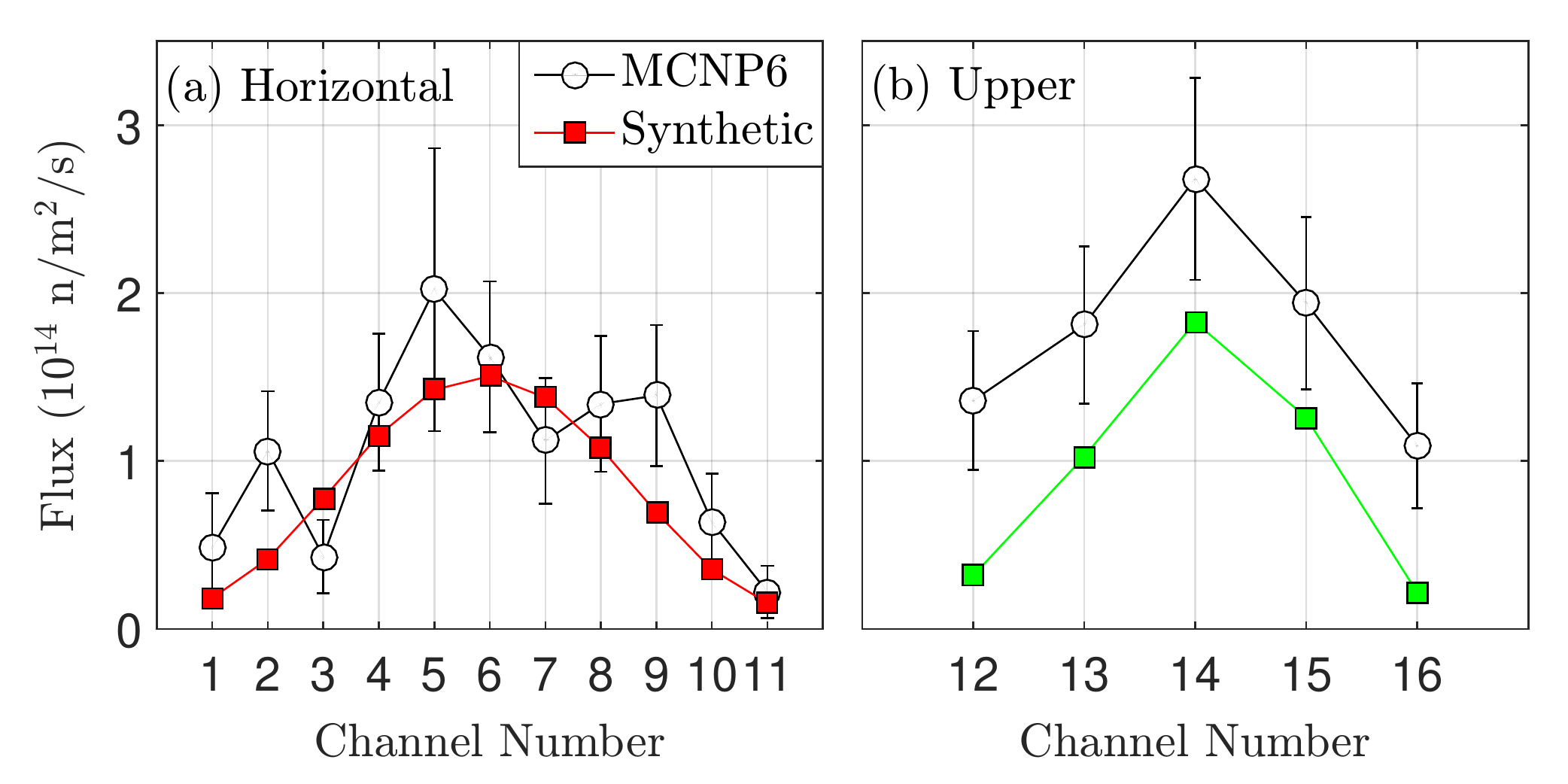}%{figures/neutCameraResults_syntheticVMCNP_180521.pdf}
  \caption{Synthetic data (squares) of line-integrated neutron flux and MCNP6 flux tallies (circles) with error bars are shown for the channels of the (a) horizontal and (b) upper cameras (see figure \ref{syntheticCamera}). A total neutron rate of $10^{19}$~n/s was assumed.}
  \label{syntheticCameraData}
\end{figure}

The proposed lines-of-sight for the MQ1 camera system are shown in figure~\ref{syntheticCamera} overlaying the DT neutron yield from figure~\ref{fig:profiles}(c). A simple calculation was performed to generate synthetic data of the neutron flux expected at the end of each collimator: At a given major radius, the number of neutrons emitted within a small volume along the line-of-sight was calculated, accounting for solid angle but no toroidal effects (i.e. a cylindrical plasma approximation was used). The flux reaching the detector was then calculated from each volume and summed to give the synthetic data for the upper and horizontal cameras shown in figure~\ref{syntheticCameraData}. As expected, the central channels are predicted to measure the highest flux as they sample the highest yield in the plasma core. The profiles are relatively symmetric as well.  

\begin{figure}[h!]
    \centering
    \includegraphics[width=\textwidth]{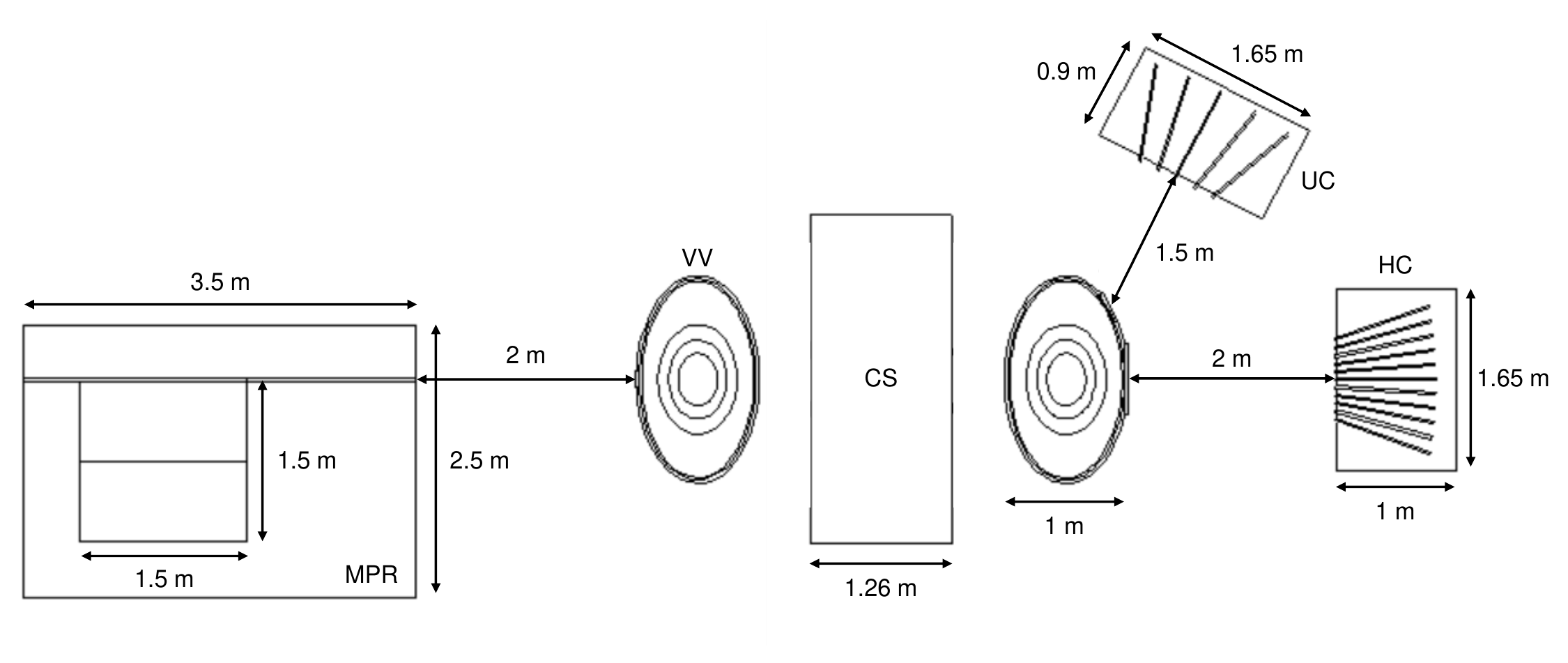}%{figures/sideView_190209.pdf}
    \caption{\red{A side view of the tokamak, radial MPR spectrometer, and horizontal and upper neutron cameras (HC/UC) modeled in MCNP6. Also labeled are the outer vacuum vessel (VV) and central solenoid (CS). Note that the tangential MPR and TF coils are not seen in this poloidal cross-section. Distances and dimensions are given, with material and other geometric details provided in the text.}}
    \label{MCNP_camerasys}
\end{figure}

The MQ1 neutron camera system was modeled in MCNP6 to better estimate neutron signal levels (including scattered neutrons) as well as background noise arising from gamma rays. The geometries of the modeled neutron camera systems \red{are} shown in figure~\ref{MCNP_camerasys}. Each camera system has a separate rectangular aperture through which it views the plasma; these were modeled as voids in the outer vacuum vessel with dimensions of 64~cm $\times$ 15~cm and 15~cm $\times$ 15~cm for the horizontal and upper cameras, respectively. (Recall that the horizontal camera aperture is the same as that for the \red{tangential} neutron spectrometer.) The horizontal camera consists of eleven collimator channels ($\sim$10~cm resolution) encased in a \red{1~m} wide, neutron-absorbing, high-density polyethylene (HDPE) box. Each collimator is 90~cm long and 1~cm in diameter. These dimensions were chosen based on the dimensions used in JET and MAST \cite{jarvis1997neutron,Cecconello2012}. The upper camera is rotated $\sim$60 degrees above the horizontal camera in the same poloidal plane and has only five collimator channels ($\sim$20~cm resolution) to minimize its size and interference with a possible upper divertor. The horizontal and upper cameras are located $\sim$2~m and \red{$\sim$1.5~m} away from their respective apertures. Given the flexibility of the system, elements like the size and number of collimators, distance between channels, type of detector assemblies, and placement of the camera system relative to the plasma can be further optimized.

In this analysis, background fluxes of down-scattered (average energy $\sim$1-2~MeV) DT neutrons were found to be at least 2$\times$ greater than higher energy neutrons; thus, an energy-resolved detector is required for useful data interpretation. These MCNP6 results incorporated the energy discrimination technique used at JET: neutrons with energies $>$7~MeV contribute to the signal, while those below the threshold are considered background \cite{jarvis1997neutron}. Figure \ref{syntheticCameraData} shows the neutron flux tallied from MCNP6 for each collimator channel, along with the synthetic data described previously. Note the similar profile shapes and absolute magnitudes. The large error bars correspond to low counting statistics arising from limited computational resources. \red{There is a noticeable discrepancy between the MCNP6 and synthetic profiles for the upper camera, as seen in figure~\ref{syntheticCameraData}b. The synthetic calculation only considers neutrons that directly enter the front face of the collimator channel. In contrast, the MCNP6 simulation also counts neutrons that scatter into the collimator body, which is a source of noise during actual operation. This effect could explain the higher fluxes and broader profile.} For both data sets, the total neutron rate was assumed to be 10$^{19}$~n/s. Typical detector efficiencies reported are $\sim$10$^{-2}$ \cite{ishikawa2002first,Adams1993}, resulting in count rates up to $\sim$10$^8$~s$^{-1}$. This allows for sufficient time resolution with good counting statistics (e.g. 1~ms and $\leq$1\% error). 

Additionally, gammas resulting from neutron interactions with the HDPE shielding can interfere with signal detection. \red{To assess this, MCNP6 simulations were performed with minimal shielding in order to determine a reference signal-to-background ratio. Results indicate that the background photon flux is almost 2-3 orders of magnitude larger than the neutron signal at the detector location. While the exact design of the detector in the neutron camera system can be optimized in future work, it is clear from this feasibility study that such a detector must be capable of both neutron energy discrimination and discrimination of gammas in order to overcome this expected poor signal-to-background ratio. Recently, the HL-2A tokamak deployed a radial neutron camera for measuring neutron emissivity profiles during DD operation \cite{zhang2016}. This design employed the Saint Gobain BC501A liquid scintillator coupled to a photomultiplier tube at the the end of each collimator channel. It was found that the use of this scintillator provided high detection efficiency and pulse-shape discrimination in order to separate gamma and neutron signals. Additional lead shielding was also used to reduce the background gamma signal. Designs similar to the HL-2A concept have also been successfully employed by JET and other machines \cite{ishikawa2002first, jarvis1997neutron} and could be used for MQ1.}

%%%%%%%%%%%%%%%%%%%%%%%%
% Summary
%%%%%%%%%%%%%%%%%%%%%%%%
\section{Summary}
\label{sec:summary}

With recent advances in high temperature superconducting technology, the high-field, compact approach to fusion has gained both interest and traction. This motivates an exploration of plasma diagnosis in the HFC parameter space\red{:} high $B$, $I_P$, $n_i$, and $T_i$, as well as compact $R$ and $a$. In this study, a suite of neutron diagnostics was presented and assessed for a conceptual SPARC-like device called the MQ1 tokamak. MCNP6 simulations of neutron transport in a notional tokamak geometry were performed to determine expected signal and noise levels. Both DD and DT fusion neutron sources were used based on physically-motivated plasma density and temperature profiles.

Three \red{widely-used and proven} neutron diagnostics were modeled to assess the feasibility of measurements supporting the three major objectives of MQ1: determining a fusion gain of $Q \geq 1$, demonstrating the use of HTS magnets, and exploring new physics of self-heating plasmas. An array of micro-fission chambers were shown to detect appropriate neutron flux levels for the calculation of fusion power and gain; however, such an array would not be appropriate for measurements of neutron emission asymmetries. The number of independent measurements required for 95\% confidence in attaining $Q \geq 1$, with expected errors, was also explored; few are needed when the measurements report $Q \geq 2$.

A magnetic proton recoil spectrometer was also assessed for energy-resolved measurements of neutron emission. Calculations of ion temperature, performed using MCNP6 energy spectral data, consistently overpredicted $T_i$ by $\sim$10-20\%. However, signal levels should be sufficiently high \red{so} that a more accurate measurement could be performed with proper calibration. Modeling also suggested that a tritium-to-deuterium fuel ratio \red{with upper limit} $n_T/n_D \approx 6\%$ is distinguishable within uncertainties. This is acceptable for trace-tritium experiments, but not for full DT operation; thus, further simulations and exploration must be done. A first look at the potential for alpha knock-on neutron measurements and tests of classical fast ion confinement indicate an improvement by a factor of $\sim$1000 compared to previous experiments. 

Finally, a two camera system was proposed for spatial neutron emissivity measurements, with the upper camera modeled off-axis to allow an upper divertor. MCNP6 simulations suggest that energy down-scattered neutrons and gamma radiation from neutron-shielding interactions could overwhelm signal levels. Therefore, both energy and pulse-shape discrimination should be employed.  MCNP6 results of energy-discriminated neutron flux at the camera detectors are similar in both magnitude and spatial profile compared to synthetic data. While an optimized design is left for future work, this work demonstrates a spatial resolution of 10-20~cm and sufficient count rates for 1-10~ms time resolution. 

%%%%%%%%%%%%%%%%%%%%%%
% Acknowledgements  
%%%%%%%%%%%%%%%%%%%%%%

\section*{Acknowledgements} 

This work originated as part of a graduate course in the Department of Nuclear Science and Engineering at the Massachusetts Institute of Technology. The authors thank Brandon Lahmann for his support in implementing the MCNP6 model; Dr. Maria Gatu-Johnson for sharing her expertise in neutron diagnostics at JET; and Drs. Daniel Brunner, Martin Greenwald, Nathan Howard, and Robert Mumgaard for useful discussion and feedback.

%%%%%%%%%%%%%%%%%%%%%%%%%%%
% Appendix 
%%%%%%%%%%%%%%%%%%%%%%%%%%%

\appendix

\section{Diagnostic operation during deuterium plasma phase}
\label{sec:DD}

While this study focused on the operation of the neutron diagnostic suite in the DT phase of MQ1 operation, we briefly explore the feasibility and additional requirements for operation during the deuterium-only plasma phase.

\subsection{Neutron flux monitors}

Most tokamaks utilizing NFMs only run deuterium plasma discharges, so this diagnostic is certainly feasible for operation during the MQ1 deuterium phase. MCNP6 simulations of deuterium plasma sources were performed for the MQ1 geometry, and assuming a total DD neutron rate of $\sim$10$^{16}$--10$^{17}$~n/s, the neutron rates impacting the MFCs are $\sim$10$^{11}$--10$^{12}$~n/s. These rates are reasonable for detection, but a longer integration time may be required to reduce counting error. Because of the difference in neutron energy between DD and DT reactions, two sets of NFMs with different amounts of shielding are likely required to measure sufficient counts. Separate DD and DT calibrations can also be performed for these diagnostics. 

Ultimately, the role of NFMs is critical during the DD phase of MQ1 operation because a ``DT fusion gain equivalent" $Q_{DT,eq}$ can be inferred from the measured DD fusion gain; that is, the DT fusion power can be predicted as if the same discharge had been run with a DT plasma, instead of a deuterium-only plasma. An indication of $Q_{DT,eq} > 1$ would motivate the MQ1 team to proceed to DT phase.

\subsection{Neutron spectrometers} 

The MPR system is capable of providing measurements during deuterium-only operation; however, the device must be optimized for it. In 2006, the MPR system on JET underwent a major upgrade allowing measurements of $T_i$, $P_{fus}$, and heating effects during deuterium operation \cite{sunden2009thin,sjostrand2006new}. In addition, a second, similarly large time-of-flight neutron spectrometer has been optimized for DD operations on JET \cite{johnson20082}. A spectrometer optimized for DD operation installed on MQ1 could aid in determining the fuel ratio during DT operation as each spectrometer could measure only one fusion neutron energy peak with higher energy resolution. However, MQ1's focused mission is designed for rapid iteration; the diagnostic suite should mirror that mission. Therefore,  minimizing the number of diagnostics will aid in MQ1's expedited development. 

\subsection{Neutron camera system}

The neutron camera design on JET was able to measure DD and DT neutron emissivity profiles simultaneously by using three detectors at the back-end of each collimator channel \cite{sasao}. A similar MQ1 design could be used during both deuterium-only and DT phases. As tritium is introduced in MQ1, the camera system could also be used to study tritium transport in the plasma and confirm previous results studied in JET \cite{bonheure2006neutron}. 

%%%%%%%%%%%%%%%%%%%%%%
% References  
%%%%%%%%%%%%%%%%%%%%%%

%\section*{References}
\bibliographystyle{unsrt}
%\bibliography{bib}

\end{document}